\documentclass[aps,prl,twocolumn,superscriptaddress,nofootinbib]{revtex4-2}

\usepackage{amsmath,amssymb}
\usepackage{graphicx}
\usepackage{bm}
\usepackage{xcolor}
\definecolor{nicegreen}{rgb}{0., 0.75, 0.46}
\definecolor{nicepurple}{RGB}{127, 38, 222}

\usepackage{hyperref}                        % must come before cleveref
\hypersetup{colorlinks=true, linkcolor=nicepurple, citecolor=nicegreen, urlcolor=nicepurple}
\usepackage[capitalize, nameinlink]{cleveref}
\crefname{endmatter}{End Matter}{End Matter}
\Crefname{endmatter}{End Matter}{End Matter}
\crefformat{endmatter}{#2End Matter#3}
\Crefformat{endmatter}{#2End Matter#3}

\newcommand{\ndata}{\bm{n}_{\mathrm{d}}}
\newcommand{\Neff}{N_{\mathrm{eff}}}
\newcommand{\dcp}{\delta_{\mathrm{CP}}}
\newcommand{\Pvac}{P_{\mathrm{vac}}}
\newcommand{\Pmat}{P_{\mathrm{mat}}}
\newcommand{\Pvacbar}{\overline{P}_{\mathrm{vac}}}

\newcommand{\numu}{\nu_\mu}
\newcommand{\nue}{\nu_e}
\newcommand{\numubar}{\bar\nu_\mu}
\newcommand{\nuebar}{\bar\nu_e}

\newcommand{\T}{T}
\newcommand{\Lmap}{\mathcal{L}}

\newcommand{\rev}[1]{{\color{teal} #1}}

\begin{document}

\title{Model-Independent Determination of Intrinsic Leptonic CP Violation}

\author{Ting Cheng}
\affiliation{Particle Theory Department, Fermi National Accelerator Laboratory, Batavia, IL 60510, USA}
\author{Kevin J.\ Kelly}
\affiliation{Department of Physics and Astronomy, Mitchell Institute for Fundamental Physics and Astronomy, Texas A\&M University, College Station, TX 77843, USA}
\author{Max Knobbe}
\affiliation{Particle Theory Department, Fermi National Accelerator Laboratory, Batavia, IL 60510, USA}
\author{Pedro A.\ N.\ Machado}
\affiliation{Particle Theory Department, Fermi National Accelerator Laboratory, Batavia, IL 60510, USA}

\date{\today}

\begin{abstract}
Establishing leptonic Charge-Parity (CP) violation is a central goal of the next generation of long-baseline neutrino oscillation experiments.
To achieve this, experimental analyses focus on excluding the CP-conserving values $\dcp=0,\pi$ within the standard three-neutrino framework.
We propose instead to establish CP violation as a phenomenon in a model-independent way.
We use Gaussian Process regression to infer the $\numu\to\nue$ and $\numubar\to\nuebar$ appearance probabilities in a parametrization-free manner at DUNE and Hyper-Kamiokande.
By combining the two experiments, with mild assumptions on how matter impacts oscillations, we estimate how they can probe intrinsic CP violation in neutrino oscillations together.
Taking the standard three-neutrino framework as a benchmark, we show that these experiments, when combined, can establish model-independent CP violation at $3\sigma$ for over 60\% of the values of $\dcp$.
\end{abstract}

\hfill{FERMILAB-PUB-26-0646-T, MI-HET-894}

\maketitle

% =====================================================================
\textbf{Introduction --}
% =====================================================================
Leptonic Charge-Parity (CP) violation connects two outstanding problems in particle physics: the origin of neutrino masses and the matter/antimatter asymmetry present in the universe today.
It has not been established experimentally, but it is a primary goal of the upcoming DUNE~\cite{DUNE:2020ypp} and Hyper-Kamiokande (HK)~\cite{Hyper-Kamiokande:2018ofw} experiments and within reach in the next decade.
These, and other long-baseline oscillation experiments~\cite{T2K:2023smv,NOvA:2025tmb,T2K:2025wet}, draw inferences regarding CP violation (CPV) by measuring $\numu\to\nue$ and $\numubar\to\nuebar$ appearance rates.
However, at a single baseline, the difference between these oscillation channels is contaminated by extrinsic CPV: the matter in Earth is not CP symmetric, inducing a CP asymmetry in neutrino propagation even when there is no intrinsic CPV.

Standard analyses separate these intrinsic and extrinsic contributions by assuming a framework for neutrino oscillations, typically the standard three-neutrino paradigm, where intrinsic CPV is governed by a phase $\dcp$ (with a matter potential that adds extrinsic contributions that violate CP).
This approach culminates in a statement about a model parameter, namely an exclusion of $\dcp\in\{0,\pi\}$, rather than a general observation of the phenomenon of CPV.
In fact, within a model which assumes unitarity, one can infer CP violation by measuring only CP-even quantities such as the moduli of 4 independent entries of the mixing matrix~\cite{Branco:1987mj, Jarlskog:1988ii, Bjorken:1988ni, Farzan:2002ct, Silva:2004gz, Botella:2005fc, Kleinknecht:2006cr, Denton:2023qmd}.

In contrast, we endeavor to genuinely observe intrinsic leptonic CPV, the difference between $\numu\to\nue$ and $\numubar\to\nuebar$ appearance rates in vacuum, that makes no assumptions about either the matter model or the oscillation framework.
In this letter, we propose a novel, model-independent method for determining intrinsic CP violation.
Model independence requires treating the probabilities $P(\numu\!\to\!\nue)$ and $P(\numubar\!\to\!\nuebar)$ not as predictions but as unknown functions of energy, to be inferred directly from data.
Gaussian-process~\cite{ wiener1949extrapolation, conti2009gaussian, williams2006gaussian, andrianakis2012effect} (GP) regression is an ideal tool for this task: rather than fixing a functional form in advance, as a polynomial or a three-flavor fit would, it assigns a probability distribution in function space and lets the data select from among them.
As a probability distribution, GPs naturally propagate statistical and systematic uncertainties to the oscillation probability.
We give the specifics of applying GP regression to the problem of neutrino oscillations in greater detail in our companion paper~\cite{Cheng:2026fwd}.

By combining reconstructed DUNE and HK probabilities, we obtain an approximate vacuum oscillation probability, under the mild assumption that matter effects enter only through the product of matter density $\rho$ and oscillation baseline $L$. 
The difference between the resulting neutrino and antineutrino vacuum probabilities is then a model-independent measure of intrinsic leptonic CP violation.
Taking the standard neutrino framework as a benchmark, this approach allows to establish CP violation in a model independent way at $3\sigma$ CL over $60\%$ of true $\dcp$ values.

% =====================================================================
\textbf{Methods \& experiments --}
% =====================================================================
We simulate experimental analyses performed at DUNE and HK and reconstruct appearance oscillation probabilities by GP regression.
DUNE is an on-axis, broad-energy beam with a $1300\,$km baseline, while HK is off-axis with a narrow-band (lower energy) beam at $295\,$km baseline.
For DUNE, we assume an exposure of $624$~kt-MW-yr, corresponding to ten years of its staged plan, split evenly between forward- and reverse-horn-current modes (FHC and RHC)~\rev{\cite{DUNE:2020ypp}}.
For HK, we take $2431$~kt-MW-yr as our reference exposure for a single-tank configuration, namely 187~kt fiducial mass, 1.3~MW beam, ten years data collection, with a 1:3 split between FHC and RHC modes. 
See~\cref{app:EndMatter} for the simulation details and input oscillation parameters.

Expected event counts in these experiments are predicted to follow a linear forward operator, $\bm{n} = A\, \bm{P} + \bm{b}$, where $\bm{n}$ represents every mode (FHC appearance and RHC appearance) event rate concatenated, $A$ represents the composition of (unoscillated) flux, cross section, detection efficiency, and energy smearing, $\bm{P}$ comprises the oscillation probabilities for all oscillatory components that enter in a mode. 
Non-oscillatory backgrounds, such as neutral current events, for each mode are written as $\bm{b}$.
In a given channel, each oscillatory component ($\nu_\mu\to\nu_e$, $\bar\nu_\mu\to\bar\nu_e$, $\nu_e\to\nu_e$ and $\bar\nu_e\to\bar\nu_e$) contributes as
\begin{equation}\label{eq:rate-component}
  \frac{dn}{dE_r} = T \sum_{\alpha}\int\! dE\,
          \frac{d\phi_\alpha}{dE}\,P_{\alpha\beta}(L, E)\,\sigma\,
          \epsilon(E_r)\,R(E_r,E)\,,
\end{equation}
where $E_r$ is the reconstructed energy, $T$ is the exposure, $\phi_\alpha$ is the flux of $\nu_\alpha$, $P_{\alpha\beta}(L, E) \equiv P(\nu_\alpha \to \nu_\beta; L, E)$, and similarly $\overline{P}_{\alpha\beta}$ for antineutrinos, $\sigma$ is the cross section, $\epsilon(E_r)$ is the detection efficiency and $R(E_r,E)$ is the energy migration matrix that maps true neutrino energy $E$ to reconstructed energy.
When building the measured spectrum, we include wrong-sign appearance events as well, which are particularly relevant for the reverse-horn-current mode, as well as disappearance of the intrinsic $\nu_e$ and $\bar\nu_e$ in the beam. 

For the model-independent approach to oscillation probabilities we use
Gaussian-process regression, where we aim to reconstruct the oscillation probabilities for all oscillatory components $\bm P$ directly from data. 
Before the GP is conditioned on data, the oscillation probabilities are governed solely by a prior $P_0 \sim \mathcal{N}(\bm{\mu}_0, K)$ (a normal distribution) where $\bm{\mu}_0$ is a constant prior mean (taken to be zero for appearance probabilities and unity for disappearance) and $K = K(\theta)$ is the prior covariance, which depends on hyperparameters $\theta$.
We will consider the dependent variable for oscillations to be $u \equiv L/E$ as it is a natural choice.
The prior correlates $P(u_i)$ and $P(u_j)$ according to how far $u_i$ and $u_j$ are from each other, but it is taken to be uncorrelated among oscillatory components, that is, $K$ is block diagonal in $P_{\mu e}$, $\overline{P}_{\mu e}$, $P_{ee}$, $\overline{P}_{ee}$. 
Here, we use the Mat\'ern-$5/2$ kernel,\!\footnote{The choice of kernel, as well as the $\nu_\mu/\bar\nu_\mu$ disappearance channels, are discussed in~\cite{Cheng:2026fwd}.}
\begin{equation}\nonumber
  K_{ij} = a^2\left(1+\sqrt{5}\,q_{ij}+\tfrac{5}{3}q_{ij}^2\right)e^{-\sqrt{5}\,q_{ij}},
  \quad q_{ij}=\frac{|u_i-u_j|}{\ell}\,.
\end{equation}
Note that $\theta = (\ell, a)$ are the hyperparameters that determine the prior.

Each channel has noise described by a Gaussian probability distribution $\mathcal{N}(\bm 0, \Sigma_n)$. 
The noise originates from statistical and systematic uncertainties and its covariance can be written as 
\begin{equation}\label{eq:noise}
  \Sigma_n = \mathrm{diag}(\bm{n}) + DD^T\,.
\end{equation}
The first term is the statistical uncertainty.\!\footnote{Technically, $\bm n$ should be the theoretical expectation, but for simplicity we use (mock) data. In the CPV test described later, $\Sigma_n$ is instead re-evaluated at the fitted expectation for the calibration data set; see~\cref{app:EndMatter}.} 
The second term describes the systematic uncertainties from nuisance parameters. 
The matrix $D$ is the response of the predicted spectrum to a change of the nuisances $\bm{\eta}$, $D_{ij}=\partial n_i/\partial \eta_j$. 
We include normalization, tilt, and energy-scale uncertainties on the forward operator and background.

The posterior distribution $p(\bm{P}|\ndata)$ gives us information on the oscillation probabilities conditioned on data. It can can written as $P\sim\mathcal{N}(\mu,\Sigma)$, where the posterior mean and covariance are 
\begin{equation}\label{eq:posterior}
  \bm{\mu} = \bm{\mu}_0 + K A^{\T}\! C^{-1} \bm{r}, \quad
  \Sigma = K - K A^{\T}\! C^{-1}\! A K\,,
\end{equation}
where $C = A K A^{\T} + \Sigma_n$ is the covariance of the counts and $\bm{r} = \ndata - A\bm{\mu}_0-\bm b$ is the residual. 

It is useful to define the the signal-to-noise ratio
\begin{equation}
  S=\Sigma_n^{-1/2}\,A K A^{\T}\,\Sigma_n^{-1/2}.
\end{equation}
Its eigenvectors are the independent modes of the measurement and its
eigenvalues $s_i$ are signal-to-noise ratios of each mode $i$.
Note that since $S$ is determined by experimental design and uncertainties alone, an experiment's information capacity can be assessed before data taking. This is typically quantified through the number of functional degrees of freedom and the mutual information
\begin{equation}\label{eq:neff}
  \Neff = \sum_i \frac{s_i}{1+s_i}\,, \qquad
  I = \frac{1}{2}\sum_i \ln(1+s_i)\,.
\end{equation}
From an information theory point of view, the mutual information can be understood as the logarithmic contraction of the posterior, relative to the prior.
That is, the posterior distinguishes $e^{I}$ times more function configurations than the prior.

The last step is how to implement the prior's hyperparameters.
In this work, we adopt a Bayesian approach where we marginalize the likelihood over the hyperparameters given some data $\ndata$,
\begin{align}\label{eq:marginalization}
  p(\bm P|\ndata)
    &= \frac{1}{Z}\int\! d\theta\, p(\bm P|\ndata,\theta)\,p(\ndata|\theta)p(\theta)\,, \\[2pt]
  \ln p(\ndata|\theta)
    &= -\tfrac{1}{2}\big[\bm{r}^{\T} C^{-1} \bm{r} + \ln\det C + {\rm const.}]\,,
\end{align}
with $Z=\int\! d\theta\, p(\ndata|\theta)p(\theta)$.
The second line is the closed-form log-likelihood, and $p(\theta)$ is the prior on the hyperparameters.
The term $\bm{r}^{\T} C^{-1} \bm{r}$ rewards hyperparameters that fit the data, while  $\ln\det C$ term favors strongly correlated covariances, i.e., small amplitude $a$ and large correlations $\ell$.
Smaller $\det C$ means the model predicts a narrower range of possible outcomes before seeing the data. 

While the probability Eq.~\eqref{eq:marginalization} is strictly not Gaussian, we can approximate it by a Gaussian distribution with mean and covariance 
\begin{align}\label{eq:marginalization_approx}
  &\hspace{-0.27cm}\langle\bm{\mu}\rangle =\frac1Z \int d\theta \,\bm{\mu}\,p(\ndata|\theta)p(\theta),\\
  &\hspace{-0.27cm}\langle\Sigma\rangle =\frac1Z \int d\theta \,\left[\Sigma\!+\!(\bm{\mu}\!-\!\langle\bm{\mu}\rangle)(\bm{\mu}\!-\!\langle\bm{\mu}\rangle)^{\T}\right] p(\ndata|\theta)p(\theta)\,,
\end{align}
where we kept implicit the $\theta$-dependence on the posterior mean $\bm{\mu}$ and the posterior covariance $\Sigma$.

We need to choose how to structure the kernels and priors on $(a,\ell)$ for each of the four probabilities $P_{\mu e}$ and $\overline{P}_{\mu e}$ at DUNE and HK, as well as for the subleading $P_{ee}$ and $\overline{P}_{ee}$.
When looking at these experiments individually, we will assume the appearance and disappearance probabilities to share the same $\ell$. 
We will also assume a shared $a$ for the appearance probabilities and we will fix $a=0.1$ for $1-P_{ee}$ and $1-\overline{P}_{ee}$, conveying our prior knowledge of the small electron-neutrino disappearance probability.
We include a mild prior on $\ell = (520 \pm 520)$~km/GeV, accounting for previous observations of oscillations associated with this scale.
For the amplitude $a$, we assume a log-uniform prior in $[10^{-2}, 1]$.
Every reported band and figure of merit below is marginalized over hyperparameters $\theta=(\ell,a)$ analogous to Eq.~\eqref{eq:marginalization_approx}.
We denote the marginalized functional degrees of freedom and mutual information by $\langle\Neff\rangle$ and $\langle I\rangle$,
which we will evaluate for DUNE and HK in the next section.

\begin{figure}[t]
  \centering
  \includegraphics[width=\columnwidth]{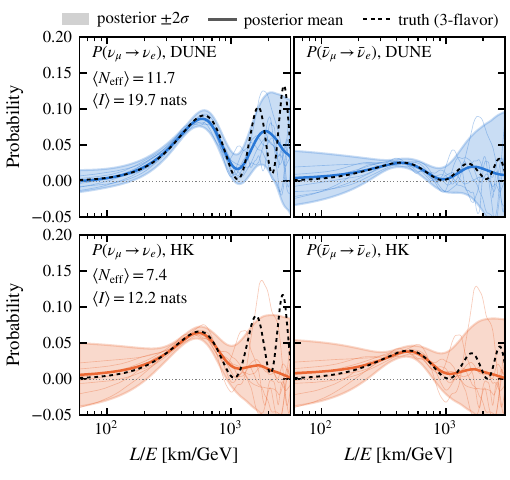}
  \caption{Model-independent reconstruction of the appearance probabilities
    $P(\numu\!\to\!\nue)$ (left) and $P(\numubar\!\to\!\nuebar)$ (right) for DUNE (top) and HK (bottom)
    as functions of $u=L/E$. We show the posterior mean and $\pm2\sigma$ band, as well as the
    three-flavor truth. 
    Thin lines are oscillation curves sampled from the GP posterior, showing examples of the functions the data support.}
  \label{fig:appearance}
\end{figure}

% =====================================================================
\textbf{Model-independent oscillations --}
% =====================================================================
The machinery of the previous section allows us to extract distributions of $P_{\mu e}$ and $\overline{P}_{\mu e}$ for DUNE and HK directly without assuming a functional form.
We show these reconstructed probabilities on a common $L/E$ axis in Fig.~\ref{fig:appearance}.
As expected, the probabilities are best determined where the event density is largest; the three-flavor curve lies within the $\pm2\sigma$ posterior band throughout, so the reconstruction is unbiased and not prior-driven.
Uncertainty propagation is automatically ensured, since the band inherits the statistical and systematic uncertainties directly through $\Sigma_n$.
All results in this letter include the detector systematics which we specify in the~\cref{app:EndMatter}.

We also show the marginalized number of functional degrees of freedom $\langle\Neff\rangle$ in each panel of Fig.~\ref{fig:appearance}.
Fitting each experiment's neutrino- and antineutrino-mode appearance channels jointly, DUNE (HK) resolves $\langle\Neff\rangle \approx 12$ (7) modes.
We have also found the marginalized mutual information $\langle I\rangle$ to be about 20  and 12 nats for DUNE and HK. 
It is interesting to note that despite larger statistics, HK has lower $\langle\Neff\rangle$ and $\langle I\rangle$ than DUNE, likely due to the narrower $L/E$ range of the neutrino spectrum.

% =====================================================================
\textbf{Intrinsic CP violation --}
% =====================================================================
Matter effects induce an extrinsic CP asymmetry in oscillations, such that even in the absence of CP violation, $P(\numu\!\to\!\nue)\neq P(\numubar\!\to\!\nuebar)$~\cite{Arafune:1997hd, Cervera:2000kp, Minakata:2001qm}. 
A single-baseline asymmetry between neutrino and antineutrino appearance channels does not, by itself, indicate intrinsic CP violation in a model-independent way. 

\begin{figure}[t]
  \centering
  \includegraphics[width=\columnwidth]{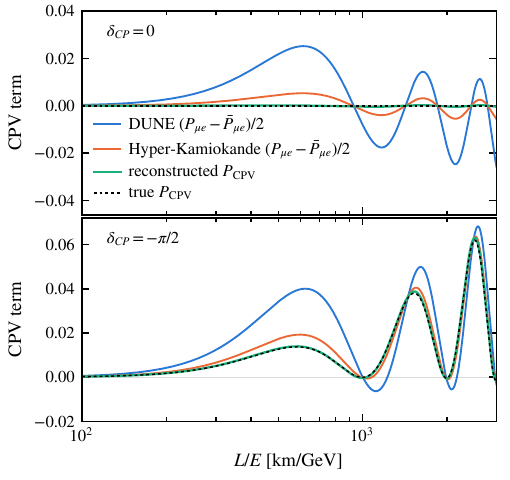}
  \caption{Comparison between CP violating term in the oscillation probability $P_{\rm CPV}$ in vacuum (dotted black) versus the approximation in Eq.~\eqref{eq:cpv_extraction} (solid green), for two values of the CP phase.
  For reference, we also show the exact differences $P_{\mu e} - \overline{P}_{\mu e}$ for DUNE (blue) and HK (orange), which are affected by matter effects.}
  \label{fig:twobaseline}
\end{figure}

Above, we extracted oscillation probabilities as a function of $L/E$, which for a given single-baseline measurement, is equivalent to measuring a probability as a function of energy.
In comparing observations across baselines with the goal of extracting the probabilities' vacuum contributions, we must make an assumption regarding the matter effects.
We assume that, beyond the dominant $L/E$-dependence, oscillation probabilities may be series-expanded in terms of the matter potential and that the leading-order term is linear in the product $\rho L$ of a given experiment's average matter density $\rho$ and baseline length $L$~\cite{Arafune:1997hd, Cervera:2000kp, Freund:2001pn, Akhmedov:2004ny,  Minakata:2015gra, Denton:2016wmg}.
Then, the oscillation probability for a given experiment ``${\rm x}$'' may be expressed as
\begin{align}
   P_{\rm x}(u) &= \Pvac(u) + \frac{\lambda_{\rm x}}{\lambda_{\rm D}}\,\Pmat(u)\,, \quad \lambda_{\rm x} = \rho_{\rm x} L_{\rm x}\,,\label{eq:split}
\end{align}
where $\Pvac$ represents the oscillation probability if the experiment were conducted in vacuum and $\Pmat$ is the deviation due to matter effects. 
For simplicity, we normalize the matter effects to the DUNE experiment, $\lambda_{\rm D} = 2.85~{\rm g/cm}^3\times1300~{\rm km}$~\cite{DUNE:2020jqi,Roe:2017zdw}, and take $\rho_{\rm H} = 2.6~{\rm g/cm}^3$ over HK's $295~{\rm km}$~\cite{Hyper-KamiokandeProto-:2015xww}. 
We assume a 3\% uncertainty on the matter density.

Both $\Pvac$ and $\Pmat$ are purely functions of $u$, and in principle are different for neutrino and antineutrino oscillations. However, we can extract the intrinsic CPV contribution to $\Pvac$ and $\Pvacbar$ by taking two experiments' oscillation measurements and constructing
\begin{align}\label{eq:cpv_extraction}
  P_{\rm CPV} &\equiv \frac{\lambda_{\rm D} \left(P_{\rm H} - \overline{P}_{\rm H}\right) - \lambda_{\rm H} \left(P_{\rm D} - \overline{P}_{\rm D}\right)}{2\left(\lambda_{\rm D} - \lambda_{\rm H}\right)}\,.
\end{align}
This combination removes anything linear in $\rho L$ and therefore is robust to a wide range of new-physics scenarios.
For example, it removes \textit{extrinsic} CPV due to non-standard matter effects, but could capture new sources of intrinsic CPV from, e.g., sterile-neutrino oscillations.

The analysis presented here does not rely on any specific parametrization of the leptonic mixing matrix and the conclusions hold
for any source of CP violation in the neutrino sector. 
Nevertheless, it is useful to show how our method performs against a concrete model. 
We thus take the standard three-flavor oscillation framework as benchmark and
show how the extraction of $P_{\rm CPV}$ works in Fig.~\ref{fig:twobaseline}: asymmetries inferred by DUNE and HK are nonzero even with $\dcp = 0$, and have similar structure even if CP is maximally violated, $\dcp = -\pi/2$, comparing blue or orange curves from one panel to another.
Taking the linear combination of Eq.~\eqref{eq:cpv_extraction} for each true $\dcp$ value yields the green lines.\!\footnote{The fact that the green lines are nearly identical to the dotted true-vacuum lines demonstrates the suitability of the $\mathcal{O}(\rho L)$ probability expansion we adopt.}
Matter effects are significant, especially for DUNE; nevertheless, this  approximation allows us to make an informed statement regarding intrinsic CPV in neutrino oscillations.
\begin{figure}[t]
  \centering
  \includegraphics[width=\columnwidth]{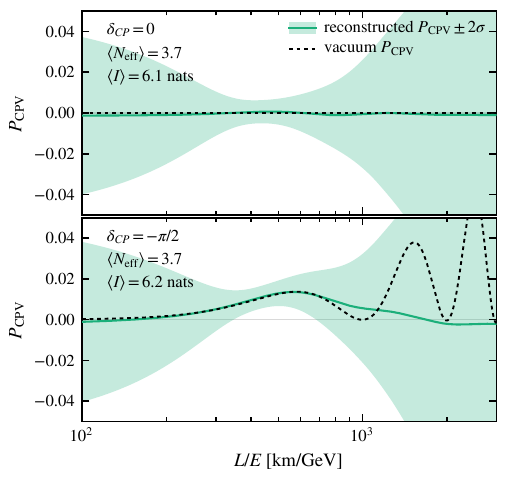}
  \caption{Intrinsic CPV term $P_{\rm CPV}$
    reconstructed from Eq.~\eqref{eq:cpv_extraction}, with the four
    appearance probabilities $P_{\rm H}$, $\overline{P}_{\rm H}$, $P_{\rm D}$,
    $\overline{P}_{\rm D}$ each fit as an independent GP (solid green, with the
    $\pm2\sigma$ posterior band in the same color), against the exact vacuum
    value (dotted black), for $\dcp=0$ and $\dcp=-\pi/2$. 
    }
  \label{fig:p_cpv}
\end{figure}
In practice, we measure this linear combination by using four independent GPs, $P_{\mu e}$ and $\overline{P}_{\mu e}$ for DUNE and for HK, on a common $L/E$ grid.
As a linear combination of Gaussian-distributed variables is also a Gaussian distributed variable, it is straightforward to build the $P_{\rm CPV}$ distribution from Eq.~\eqref{eq:cpv_extraction} by determining its mean and covariance matrix.
This determination is shown in Fig.~\ref{fig:p_cpv}, assuming the standard oscillation framework as truth, for two different values of $\dcp = 0$ and~$-\pi/2$.
We show the reconstructed $P_{\rm CPV}$ obtained from Eq.~\eqref{eq:cpv_extraction}, together with the $\pm2\sigma$ uncertainty band.
With these exposures and for the extreme $\dcp = -\pi/2$ case, intrinsic CPV can, in fact, be determined in a model independent way.
\begin{figure}[t]
  \centering
  \includegraphics[width=\columnwidth]{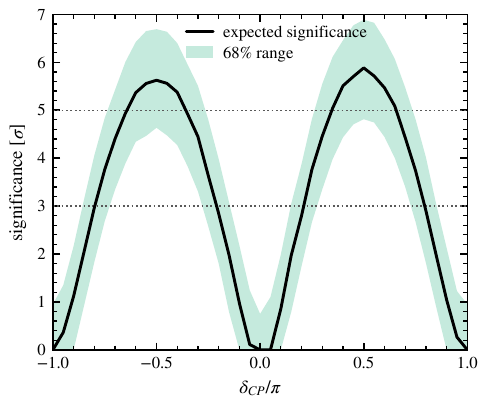}
  \caption{Expected significance of the model-independent CP-violation test for the standard three-neutrino oscillation benchmark 
    as a function of $\dcp$.
    The black curve is the median over pseudo-experiments, the band is
    their $68\%$ spread, see~\cref{app:EndMatter} for details.
    }
  \label{fig:cpv_sensitivity}
\end{figure}

Finally, we quantify this sensitivity in Fig.~\ref{fig:cpv_sensitivity}, showing the capability of DUNE and HK to jointly observe intrinsic CPV, again using the standard three-neutrino framework as a benchmark.
The details of the test statistics used and its calibration can be found in the~\cref{app:EndMatter}.
For $60\%$ of true $\dcp$ values, our model-independent approach can identify intrinsic CPV at $3\sigma$~CL or higher, and for $30\%$ of values at $5\sigma$~CL or higher.
Note that our model-independent determination of intrinsic CP violation is a much harder question to answer than the usual analysis of CP violation within a three-flavor framework.
Moreover, measurements of $P_{\mu e}$ and $\overline{P}_{\mu e}$ from multiple baselines -- DUNE \textit{and} HK -- are necessary to attain this ambitious goal.

% =====================================================================
\textbf{Conclusion --}
% =====================================================================
We have presented a method to establish leptonic CP violation as a phenomenon, rather than a model-specific question such as the exclusion of $\dcp\in\{0,\pi\}$.
We treat the appearance probabilities of DUNE and HK as free functions of $L/E$, reconstructed by Gaussian-process regression.
Combining the two baselines through the ratio of their $\rho L$ values cancels the matter contribution to first order, including any non-standard component, and isolates the intrinsic CP-odd probability $P_{\rm CPV}$.
Taking standard three-neutrino oscillations as an example of the truth, we have shown that DUNE and HK together can establish intrinsic CP violation in a model independent way using our proposal at 
$3\sigma$ for $60\%$ of the range of $\dcp$ and at 5$\sigma$  for 30\% of the $\dcp$ values.
Our method could also give access to the CP-conserving and matter-effect probabilities through other linear combinations, similar to Eq.~\eqref{eq:cpv_extraction}, which we leave to future work.

\begin{acknowledgments}
\noindent\textit{Acknowledgments} ---
This manuscript is based upon work supported by Fermi Forward Discovery Group, LLC under Contract No. 89243024CSC000002 with the U.S. Department of Energy, Office of Science, Office of High Energy Physics.
The work of P.M.\ and M.K.\ was supported by the U.S. Department of Energy,
Office of Science, Office of Advanced Scientific Computing Research,
Scientific Discovery through Advanced Computing (SciDAC-5) program,
grant ``NeuCol'', DE-AC02-07CH11359.
K.J.K.\ was supported in part by Department of Energy Award No.\ DESC0010813.
K.J.K.\ thanks the Fermilab Theory Division for their hospitality and support through the Summer Visitors' Program during a portion of this work.
During this work, we used Claude (Anthropic) to assist with writing code and generating figures. All code and visualizations were reviewed and verified and we take full responsibility for the results.
\end{acknowledgments}

\appendix
\onecolumngrid
\section*{End Matter}
\phantomsection
\makeatletter\def\@currentlabel{}\def\cref@currentlabel{[endmatter][][]}\makeatother
\label{app:EndMatter}
\twocolumngrid
\setcounter{equation}{0}
\renewcommand{\theequation}{E\arabic{equation}}
\setcounter{figure}{0}
\renewcommand{\thefigure}{E\arabic{figure}}

\textbf{DUNE and HK Simulation Details --}
All results take the NuFit~6.1 normal-ordering benchmark as the true oscillation parameters (except for varying $\dcp$).
DUNE is simulated at 624~kt-MW-yr, the ten-year staged exposure of its design report~\cite{DUNE:2020ypp}, split equally between neutrino and antineutrino running.
For systematics we take independent normalizations and spectral tilts of the signal and background components of each mode ($2\%$ signal and $5\%$ background normalizations, $5\%$ tilts), plus a $2\%$ energy scale applied as a count-conserving remap of the reconstructed spectrum.

For HK, we assume a single tank with 2431~kt-MW-yr (187~kt fiducial mass, 1.3~MW, ten years~\cite{Hyper-Kamiokande:2018ofw}), split 1:3 between FHC and RHC.
The HK appearance response includes the non-quasielastic contribution following Ref.~\cite{Machado:2013kya}, with the sample composition matched to Ref.~\cite{Hyper-Kamiokande:2025fci}.
We adopt the ``improved'' systematics budget published by the collaboration~\cite{Hyper-Kamiokande:2025fci}.
Each single-ring $e$-like sample carries a normalization uncertainty ($2.1\%$ and $2.2\%$ for the FHC- and RHC-modes), and the two modes are correlated with correlation coefficient $0.64$.
Ref.~\cite{Hyper-Kamiokande:2025fci} does not give information on shape uncertainties.
We take a spectral tilt of  $2\%$, and an energy scale uncertainty of $0.17\%$, both set to reproduce the published $\dcp$ resolution.

We do not assume any correlated systematics between DUNE and HK and we validated our simulations against the official predictions from DUNE and HK.

\textbf{Quantifying Significance of CPV --}
The model-independent test compares the reconstructed function $P_{\rm CPV}$ of Eq.~\eqref{eq:cpv_extraction} against a null hypothesis $H_0$ where we assume $P_{\rm CPV}^{\rm test}(u)=f(u)$; we formulate the test for a general $f$ and set $f=0$, the absence of CP violation, only at the end.
The null hypothesis is obtained by conditioning the GP prior on $\Lmap\bm{P}=\bm{f}$, where $\bm{P}$ collects the four appearance probabilities, and $\Lmap$ is the linear map of Eq.~\eqref{eq:cpv_extraction}.
This leads to the (conditioned) prior mean and covariance
\begin{align}\label{eq:null_prior}
  \bm{\mu}_{0,H_0} &= \bm{\mu}_0 + K \Lmap^{\T}\left(\Lmap K \Lmap^{\T}\right)^{-1} \left(\bm{f}-\Lmap\bm{\mu}_0\right),\\
  K_{H_0} &= K - K \Lmap^{\T}\left(\Lmap K \Lmap^{\T}\right)^{-1} \Lmap K\,,\nonumber
\end{align}
where $\bm{\mu}_0$ and $K$ are the GP prior mean and covariance.
For our specific case, we have set $\bm{\mu}_0=0$.
By construction, every function drawn from this prior has $P_{\rm CPV}(u)=f(u)$, while the CP-even and matter-induced parts of the four probabilities keep the original GP prior.

The two hypotheses are compared through a test statistic built from the ratio of their evidences, i.e.\ their marginal likelihoods,
\begin{align}\label{eq:q}
  q &= 2\Big[\ln Z  - \max_{s}\ln Z_{H_0}(s)\Big]\,,\\
  Z & = \int\! d\theta\,\mathcal{N}\!\left(\bm{r};\,\bm{0},\,AKA^{\T}+\Sigma_n\right) p(\theta)\,,\nonumber
\end{align}
where $p(\theta)$ is the prior on hyperparameters $\theta$ and $\bm{r}= \bm{n} - A\bm{\mu}_0-\bm b$ is the residual between prior mean and the analyzed dataset $\bm{n}$.
We write $q_{\rm d}$ for $q$ evaluated on the observed data, $\bm{n}=\ndata$. 
The null evidence $Z_{H_0}(s)$ is defined in the same way as $Z$, but with $K\to K_{H_0}$ and $\mu_0\to\mu_{0,H_0}$. 
Hence, $q$ measures how the evidences differ after removing the CP-odd freedom from the prior. 
In addition, to account for uncertainties on the matter density, we maximize over $s\in\{0.97,\,1,\,1.03\}$ which multiplies the ratio $\lambda_{\rm D}/\lambda_{\rm H}$.

While our statistic has the form of the likelihood ratio of Ref.~\cite{Cowan:2010js}, the probabilities and the nuisance parameters are marginalized over their priors (there, they are profiled), and thus $q$ does not necessarily follow a $\chi^2$ distribution. 
Therefore, the significance of $q_{\rm d}$ needs to be calibrated through its $p$-value under the distribution of $q$ for CP conserving data. 
We generate this distribution through pseudo-experiments that mock CP-conserving data model independently, replacing $\ndata$ in Eq.~\eqref{eq:q} with pseudo-data generated around a CP conserving spectrum $\bm{n}_{H_0}$. 

However, there is one challenge in performing this calibration.
In a data-driven analysis, the noise model and the fit depend on each other: the Poisson variance in $\Sigma_n$ and the detector response $D$ should be evaluated at the expected spectrum $\bm{n}_{H_0}$, while $\bm{n}_{H_0}$ is itself the outcome of a fit that uses $\Sigma_n$.
To circumvent this issue, the spectrum $\bm{n}_{H_0}$ is determined in the following way. 

We start with $\Sigma_n={\rm diag}(\ndata)+DD^T$ (evaluated at the data); then we replace $K\to K_{H_0}$ in Eq.~\eqref{eq:posterior}, which yields the posterior mean $\bm{\mu}_{H_0}$ and pulls $\bm{\eta}_{H_0}$
for each hyperparameter $\theta$. 
Next, we marginalize 
\begin{align}\label{eq:null_marg}
  \langle\bm{\mu}_{H_0}\rangle &=\frac1{Z_{H_0}} \int d\theta \,\bm{\mu}_{H_0}\,p(\ndata|\theta,H_0)\,p(\theta)\,,\nonumber\\
  \langle\bm{\eta}_{H_0}\rangle &=\frac1{Z_{H_0}} \int d\theta \,\bm{\eta}_{H_0}\,p(\ndata|\theta,H_0)\,p(\theta)\,.
\end{align}
Note that $\Lmap \langle\bm{\mu}_{H_0}\rangle=\bm{0}$ by construction. 
With the forward operator and background model fixed at the best fit $\langle\bm{\eta}_{H_0}\rangle$, we obtain the data-driven null hypothesis spectrum and the systematic response 
\begin{equation}
  \bm{n}_{H_0}= A_{H_0} \langle\bm{\mu}_{H_0}\rangle + b_{H_0},\quad D_{H_0}=\partial\bm{n}/\partial\bm{\eta}\,.
\end{equation}  
Note that in general $\bm{n}_{H_0}\neq \ndata$.
We then replace $\Sigma_n\to \mathrm{diag}(\bm{n}_{H_0}) + D_{H_0}D_{H_0}^{\T}$ and repeat the procedure to update $\bm{n}_{H_0}$ until it converges. 

With the (converged) quantities $\bm{n}_{H_0}$, $\langle\bm{\eta}_{H_0}\rangle$ and $D_{H_0}$, we generate pseudo-experiments following the convention of the LHC Higgs combination~\cite{ATLAS:2011tau}: every pseudo-experiment draws $\bm{n}_{\rm toy}$ from a Poisson distribution with mean $\bm{n}_{H_0}$, and draws $\bm{\eta}_{\rm toy}$ from $\mathcal{N}(\langle\bm{\eta}_{H_0}\rangle,I)$. 
Each mock dataset is analyzed as the data, with residual $\bm{r}_{\rm toy} = \bm{n}_{\rm toy} - A\bm{\mu}_0 - \bm{b} - D_{H_0}\bm{\eta}_{\rm toy}$, resulting in a test statistic $q$. 
The significance of CPV is then determined by the usual fraction of pseudo-experiments with $q>q_{\rm d}$. 

To obtain the expected significance in Fig.~\ref{fig:cpv_sensitivity}, we follow the procedure described above for the three-flavor benchmark for each input value of $\dcp$.
Each of them yields its own $q$, which fluctuates with the statistical and systematic uncertainties of $\ndata$. Figure~\ref{fig:cpv_sensitivity} shows the median sensitivity $Z$ and its $1\sigma$ range.
As building one null distribution per pseudo-experiment would be prohibitive, we have checked that the shape of the null distribution of $q$ is the same across $\dcp$ to the precision required, and we refer all pseudo-experiments to a single null distribution, generated at the expected data for $\dcp=-\pi/2$.
For $p$-values that are too small, we obtain the tail of the null distribution by importance sampling. 

\bibliography{literature.bib}

\end{document}